# TSALLIS´ ENTROPY: A NON-EXTENSIVE FREQUENCY-MAGNITUDE DISTRIBUTION OF EARTHQUAKES.

**Oscar Sotolongo-Costa [1,2,4]; Antonio Posadas[2,3,4]**

(1) Department of Theoretical Physics, University of Havana, 10400 Havana, Cuba.

(2) Department of Applied Physics, University of Almería, 04120 Almería, Spain.

(3) Andalusian Institute of Geophysics and Seismic Disasters Prevention, University of Almería, 04120 Almería, Spain.

(4) "Henri Poincaré" Chair of Complex Systems, University of Havana, 10400 Havana, Cuba.

## ABSTRACT

By using the maximum entropy principle, with Tsallis´ entropy, we obtain an explicit dependence for energy distribution of earthquakes. This function describes very well the observations in a wide range of energies, where other distribution functions fail. We assume that the fragments filling the gap between the fault planes play an active role in the triggering of earthquakes. The energy distribution function is related to the size distribution function of these fragments.

## INTRODUCTION

A great number of papers have been originated as a result of the Gutenberg Richter law, where the importance of the knowledge of the energy distribution of the earthquakes and its physical and practical implications are emphasized. Some famous models, like those of Burridge and Knopoff (1967) or Olami et al. (1992), have focused on the mechanical phenomenology of earthquakes through simple images which capture essential aspects of the nature and genesis of a seism; these include the relative displacement of tectonic plates or the relative motion of the hanging wall and footwall on a fault, as well as the existence of a threshold for a catastrophic release of energy in the system.

Today it is widely accepted that most earthquakes have their origin in the relative motion of fault planes, whereas the images on which this energy release is modelled are diverse. The standard picture usually assigns the cause of an earthquake to some kind of rupture or to a stick-slip mechanism in which the friction properties of the fault play the determinant role. A review of these viewpoints and a few generated paradoxes can be found in Sornette (1999).

The influence of the fault profiles and the size distribution of the fragments filling the gap between the blocks of the fault in the characteristics of earthquakes have been highlighted; for example, the irregular geometry of the profiles of the tectonic plates and fault planes was highlighted in De Rubeis et al. (1996) using a geometric viewpoint to obtain the power law dependence of the earthquake energy distribution with good results.

In addition, the importance of a geometric viewpoint to study the phenomenon of fault slip has also been treated in Herrmann et al. (1990), where an idealized representation of the fragmented core of a fault (gouge) is presented. Herrmann et al. (1990) presents the gouge as a medium formed by circular disk-shaped pieces which act like bearings filling the space between two planes.

In this paper, we present a more realistic approximation by considering that the surfaces of the tectonic plates are irregular and that the space between them contains fragments of a diverse shape. We will present the "geometric" image which involves the fragments and irregularities between the two plates with a fragment size distribution deduced from a non-extensive formulation by maximizing Tsallis´ entropy. We assume that the physical mechanism for the triggering of an earthquake suggests a relation between the fragment size distribution and the energy distribution of earthquakes, as it will be explained later.

The Gutenberg-Richter law expresses the log-linear dependence between the number of earthquakes of a magnitude greater than a given one and the value of this magnitude. However, the graphical representation of this law for different catalogues reflects that for the smallest magnitudes the dependence is not fulfilled. It is usual to consider that this misalignment is due to the threshold of sensitivity of the instruments and therefore, the catalogue is complete up till the value of the minimum magnitude for which the Gutenberg-Richter relation is fulfilled. In this paper we considered that before arriving to the threshold value of sensitivity of the instruments, the curvature that exhibits the frequency-magnitude relation (fig.1) can be explained assigning a fundamental role to the existence of the fragments between the planes of the fracture.

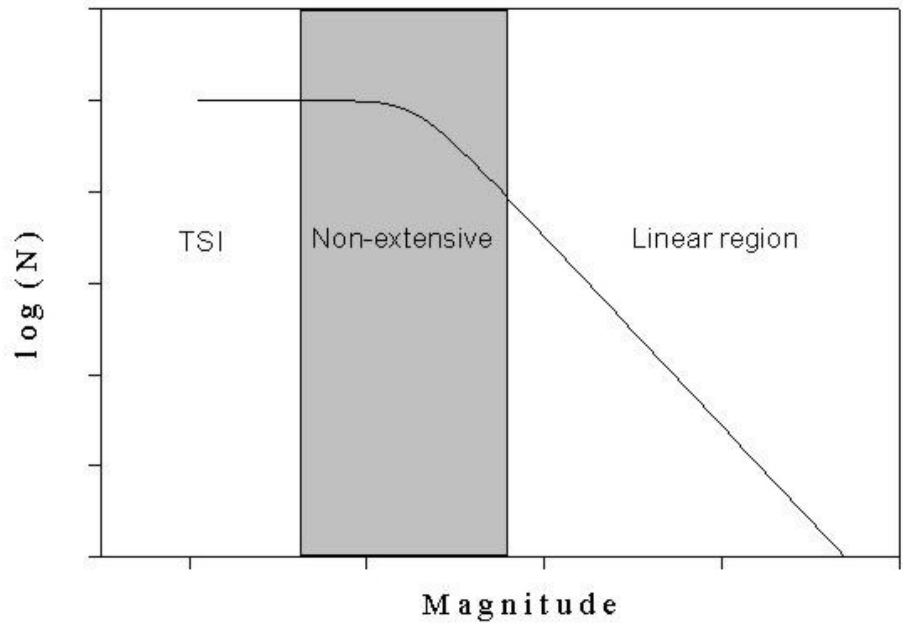

*Figure 1. Typical draw for the frequency-magnitude relation. Three areas can be considered: the threshold sensibility for instruments (TSI) region, the non-extensive area and the linear area.*

For large magnitudes, however, the Gutenberg-Richter law also fails, revealing thus the limitations of this empirical formula, while the model we present here describes very well the energy distribution all through the range of magnitudes.

We will compare our theoretical results with small earthquakes registered in the south of the Iberian Peninsula (Spain) and with large earthquakes reported in Lomnitz and Lomnitz (1979). Afterwards we will apply our function to a whole catalogue in two cases: firstly to earthquakes in California and then to those in the Iberian Peninsula (Spain).

**IMAGE FOR EARTHQUAKES**

The irregularity of the borders of the tectonic plates has been pointed out as a main source of earthquakes and in De Rubeis et al. (1996) the Gutenberg-Richter law was obtained from computer simulations through assuming a brownian shape of the profiles and with the hypothesis that the energy release is proportional to the overlap interval between profiles.

In other models (Herrmann et al., 1990) the material between the fault planes is considered; in this case, as we already pointed out in the introduction, an ideal collection of spheres of different sizes between two plane surfaces is studied. This image can be applied, among other things, to the explanation of the eventual displacement of tectonic plates without the occurrence of a seism, since in this case the spheres would act as roll bearings.

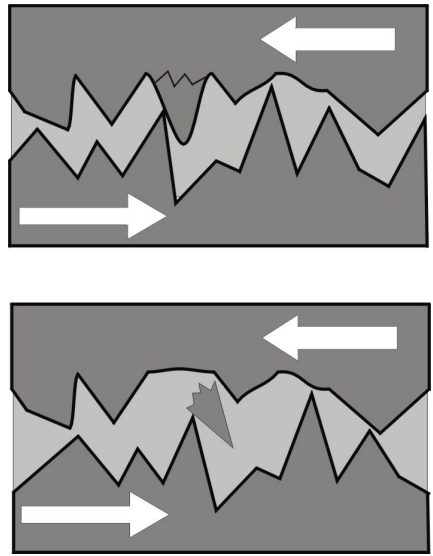

*Figure 2. An illustration of the relative motion of two irregular faults when an asperity or a barrier is broken.*

Nevertheless, these images can inspire another; i.e., that the irregularities of the fault planes can be combined with the distribution of fragments between them to develop a mechanism for triggering earthquakes; then, it is tempting (see Saleur et al., 1996) to relate fragment size distribution function with the energy distribution of the earthquakes.

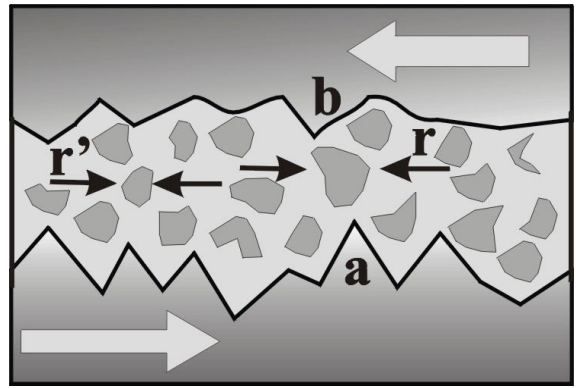

*Figure 3. An illustration of the relative motion of the planes of a fault with material between them. This material may play the role of bearings and also of particles that hinder the relative motion of the planes, as it is shown between the points a and b in the figure.*

To start, let us consider the situation illustrated in figure 2, as proposed by De Rubeis et al. (1996): two irregular profiles are able to slip. Stress in the structure accumulates until one of the asperities is broken; then, the slip occurs. But, on the other hand, we can consider the phenomenon as shown in figure 3. The motion can be hindered not only by the overlapping of two asperities of the profiles, but also by the eventual relative position of several fragments between two points "*a*" and "*b*". Stress in the resulting structure accumulates until a displacement of one of the asperities, due to the displacement of the hindering fragment, or even its breakage at the point of contact with the fragment, leads to a relative displacement of the fault planes of the order of the size "*r*" of the hindering fragment.

It is natural to think that the displacement of fragments is  more frequent than the breakage of asperities, and so most of the earthquakes (though not all of them) may have their origin in that mechanism. The eventual release of stress, whatever be the cause, leads to a displacement with the subsequent liberation of energy. We assume this energy "ε" to be proportional to "*r*",  and so the energy distribution of the earthquakes generated by this mechanism can reflect the size distribution of the fragments in the gouge.

**THE MODEL**

As already pointed out, the size distribution function of fragments between the fault planes can be expressed through the energy distribution function of earthquakes.

We can assume that the fragments are the result of the local breakage due to the constant interaction of the fault planes . The process of fault slip can be considered to occur in a homogeneous fashion through out the depth of the fault so that in any plane transverse to the fault the situation is the same. To deduce the size distribution function of the fragments we consider a two-dimensional frame as the one illustrated in figures 2 or 3. Our problem is to find the distribution of fragments by area.

To do this, we will apply a very general principle of physics: the maximum entropy principle, in the same way as we did in a previous paper (Sotolongo-Costa et al, 2000).

The Boltzmann-Gibbs formulation in the maximum entropy principle proved to be useful in the study of the fragmentation phenomena realized by Englman et al. (1987); but in this study an important feature of the fragmentation, i.e. the eventual presence of scaling in the size distribution of fragments, was not obtained and the size distribution function obtained does not fit in with all the experimental results.

The process of violent fractioning of the fault planes, producing the fragments between them, leads to the existence of long range interactions among all the existent fragments. Fractioning is then a paradigm of non-extensivity. This suggests that it may be necessary to use non-extensive statistics, instead of the one of Boltzmann-Gibbs, to describe the size distribution function of the fragments.

We will apply the maximum entropy principle with the Tsallis entropy (Tsallis, 1988) and compare the results with those obtained using the Boltzmann entropy. The Tsallis entropy for our problem has the form:

$$S_q \quad = \quad k \quad \frac{1 - \int p^q(\sigma) d\sigma}{q - 1} \tag{1}$$

where $p(\sigma)$ is the probability of finding a fragment of relative surface $\sigma$ referred to a characteristic surface of the system, and $q$ is a real number. $k$ is Boltzmann's constant. It is easy to see that this entropy is the Boltzmann entropy when $q \rightarrow 1$. The sum in all states in the entropy is here expressed through the integration over all the sizes of the fragments.

The maximum entropy formulation for Tsallis´ entropy involves the introduction of at least two constraints. The first one is the normalization of $p(\sigma)$:

$$\int\limits_{0}^{\infty} p(\sigma)d\sigma = 1 \qquad (2)$$

and the other is the "*ad hoc*" condition about the $q$-mean value, which in our case can be expressed as:

$$\int\limits_{0}^{\infty} \sigma p^{q}(\sigma)d\sigma = << \sigma >>_{q} \qquad (3)$$

This condition reduces to the definition of the mean value when $q \rightarrow 1$. More information concerning the constraints that can be imposed in the formulation can be seen in Tsallis et al. (1998). This formulation of the Statistical Physics, known as "non extensive" formulation, since this entropy is not additive, proved to be very useful in describing phenomena in which Boltzmann´s statistics fails to give a correct explanation, especially when the spatial correlations cease to be short ranged (Tsallis, 1999).

As we have already said, fracture is a paradigm of such long-range interaction phenomenon, and we gave a formulation in terms of Tsallis statistics very recently with results that explain the experimental behavior of fragmentation phenomena (Sotolongo-Costa et al., 2000). Then, the problem is to find the extremum of $\dfrac{S_q}{k}$ subject to the conditions given by formulae 2 and 3. To simplify this we will assume $<<\sigma>>_q = 1$; we will see that this has no effect on the final result.

To apply the method of Lagrange multipliers we define the lagrangian function $\Gamma$ as:

$$\Gamma = \frac{S_q}{k} + \lambda \int_0^\infty p(\sigma)d\sigma + \beta \int_0^\infty \sigma p^q(\sigma)d\sigma \qquad (4)$$

being $\lambda$ and $\beta$ the Lagrange multipliers. Application of the method follows with:

$$\frac{\partial \Gamma}{\partial p} = 0 \qquad (5)$$

and with the application of the conditions 2 and 3. So, it is possible to find:

$$p(\sigma)d\sigma = \frac{(2-q)^{\frac{1}{2-q}} d\sigma}{\left[1 + (q-1)(2-q)^{\frac{q-1}{2-q}}\sigma\right]^{\frac{1}{q-1}}} \qquad (6)$$

for the area distribution of the fragments of the fault plates.

If we now introduce that the released relative energy $\varepsilon$ is proportional to the linear dimension $r$ of the fragments, as $\sigma$ scales with $r^2$, the resulting expression for the energy distribution function of the earthquakes due to this mechanism is:

$$p(\varepsilon)\,d\varepsilon = \frac{2C_1 k\varepsilon\,d\varepsilon}{\left[1 + C_2 k\,\varepsilon^2\right]^{1/(q-1)}} \qquad (7)$$

with $C_1 = (2\text{-}q)^{1/(2\text{-}q)}$ and $C_2 = (q\text{-}1)(2\text{-}q)^{(q\text{-}1)/(2\text{-}q)}$ and the probability of the energy of an earthquake is $p(\varepsilon) = n(\varepsilon)/N$ being $n(\varepsilon)$ the number of earthquakes of energy $\varepsilon$ and $N$ the total number of earthquakes; $k$ is the proportionality constant between $\sigma$ and $\varepsilon$.

Hence, we have obtained an analytic expression which describes the energy distribution of earthquakes. This was obtained from a simple model starting from first principles. No *ad hoc* hypothesis was introduced but the proportionality of "$\varepsilon$" and "$r$", which seems justified (of course, a similar treatment can be performed with Boltzmann's entropy).

To use the common frequency-magnitude distribution, the cumulative number N($>\varepsilon$) of earthquakes with energy greater than $\varepsilon$ was calculated as the integral from "$\varepsilon$" to "$\infty$" of the formula 7; then:

$$\frac{N(>\varepsilon)}{N} = \int_{\varepsilon}^{\infty} p(\varepsilon)d\varepsilon \qquad (8)$$

where $N$ is the total number of earthquakes. On the other hand $m \propto \log(\varepsilon)$ where $m$ is the magnitude, so we get:

$$\log\left(\,N(>m)\,\right) = \log N + \left(\frac{2-q}{1-q}\right)\log\left[1 + k(q-1)(2-q)^{\frac{1-q}{q-2}} \cdot 10^{2m}\right] \qquad (9)$$

This is not a trivial result, and incorporates the characteristics of non-extensivity into the distribution of earthquakes by magnitude. Whereas the use of Boltzmann's entropy with the same method leads to:

$$\log N(>m) = a - b.10^{2m} \qquad (10)$$

with $a$ and $b$ two constants to be adjusted with the data.

## DATA AND APPLICATION OF THE MODEL

### Test for small earthquakes

The Andalusian Institute of Geophysics and Seismic Disasters Prevention compiled the earthquake catalogue used in this study. The Andalusian Seismic Network consists of more than 20 observational stations (Posadas et al., 2000). The analyzed area is the region between 35º and 38º north latitude and between 0º and 5º west longitude. The catalogue is comprised of more than 20000 earthquakes. The errors of the hypocenters location in the *x*, *y* and *z* directions are about ±1 *km*, ±1 *km* and ±2 *km*, respectively (Posadas et al., 1993). The seismicity during the period 1985-2000 may be considered normal, i.e. without major seismic events.

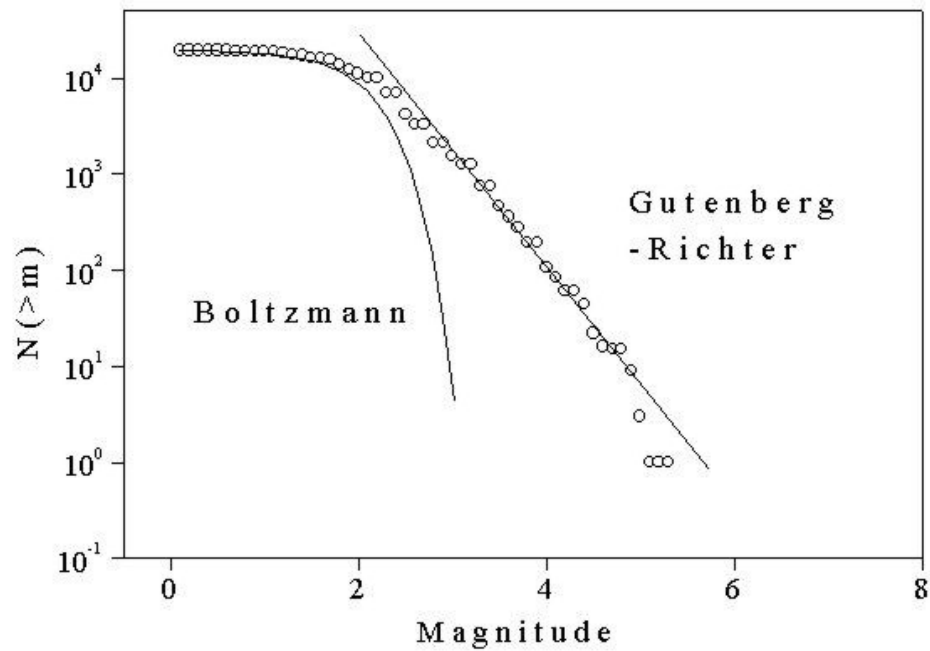

*Figure 4. Bolztmann's description (formula 10) and the classical Gutenberg-Richter's fit for the south of the Iberian Peninsula (Spain). Formula 10 does not describe the earthquakes with magnitude higher than 2. Gutenberg-Richter´s law reflects the power-law distribution for moderate seismicity.*

The Gutenberg-Richter relation is satisfied in this data set, for earthquakes with magnitude greater than 2.5. The data is assumed to be free of observational bias as well as of abnormal seismicity. Boltzmann's description with formula 10  and Gutenberg-Richter's fit are in figure 4 whereas the description with formula 9 based in a non extensive formulation is shown  in figure 5.

The assumptions applied to the Boltzmann formulation do not work for earthquakes of large or even moderate magnitude. The frequency of event occurrence, as we expected, works correctly only for low seismicity in the region.

Finally, Tsallis's formulation helps to show that our assumptions are correct because the curve adjusts itself very well for seismicity ranging from 0.0 to 2.5 magnitude; formula 9 gives also for higher magnitude a good agreement for all the magnitude values. The obtained *q* value is 1.65 ± 0.01 with a correlation factor equal to 0.99885.

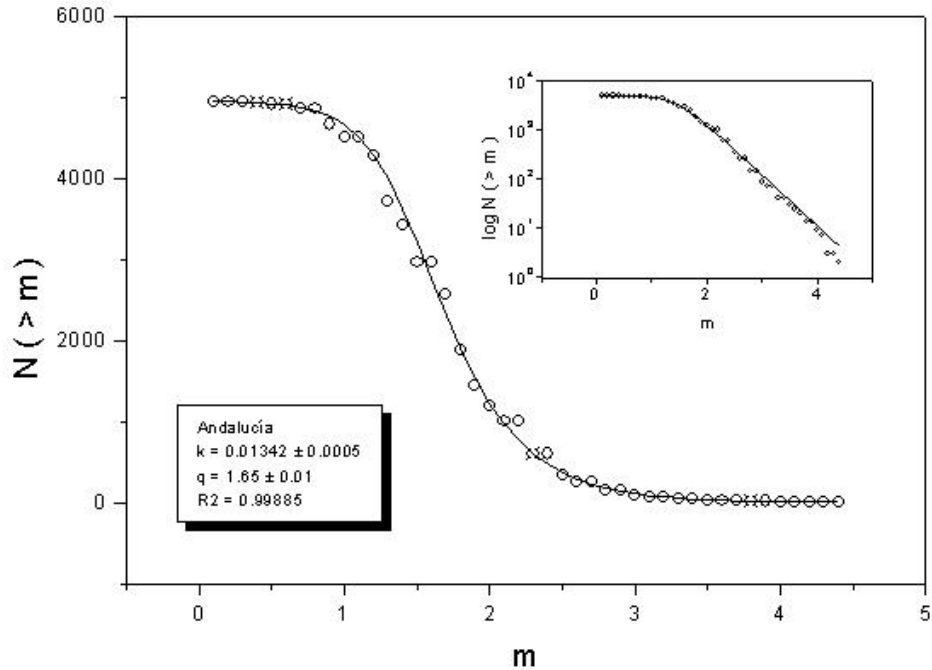

*Figure 5. Non extensive description (formula 9) for the south of the Iberian Peninsula. Our model points out that the fit is possible for earthquakes ranging from 0.0 to 2.5 of magnitude. The correlation factor is 0.998 and the value of q is 1.650. In the upper right corner, the usual image of the frequency-magnitude relation by using log scale for the number of earthquakes is shown.*

**Test for large earthquakes**

Lomnitz and Lomnitz (1979) have proposed a stochastic model of strain accumulation and release at plate boundaries. The model leads to a generalized Gutenberg-Richter´s relation in terms of *G(m),* the cumulative excedence of a magnitude *m*, which tends to the original one of Gutenberg-Richter in the low magnitude range and which provides estimates of the probability of occurrence, significantly more adequate than the Gutenberg-Richter law, at high magnitudes. They have obtained an excellent agreement with the data of the Chinese earthquake catalogue, which contains the earthquakes for a threshold magnitude *m = 6.0*; it is the longest published catalogue of historical earthquakes in any region. Lomnitz and Lomnitz (1979) excedence is defined as:

$$G(\varepsilon) = \int\limits_{\varepsilon}^{\infty} p(\varepsilon)d\varepsilon \qquad (11)$$

Using this definition with our formula (7) and expressing the result with magnitude instead of energy, we obtain:

$$\log\left(\,G(>m)\,\right) = \left(\frac{(2-q)}{1-q}\right)\log\left[1+k(q-1)(2-q)^{1-q/q-2}\cdot 10^{2m}\right] \qquad (12)$$

We adjusted the constants of this equation with the Chinese catalogue and the results can be seen in figure 6; in this case $q = 1.6877 \pm 0.0001$ and the correlation factor is 0.9925.

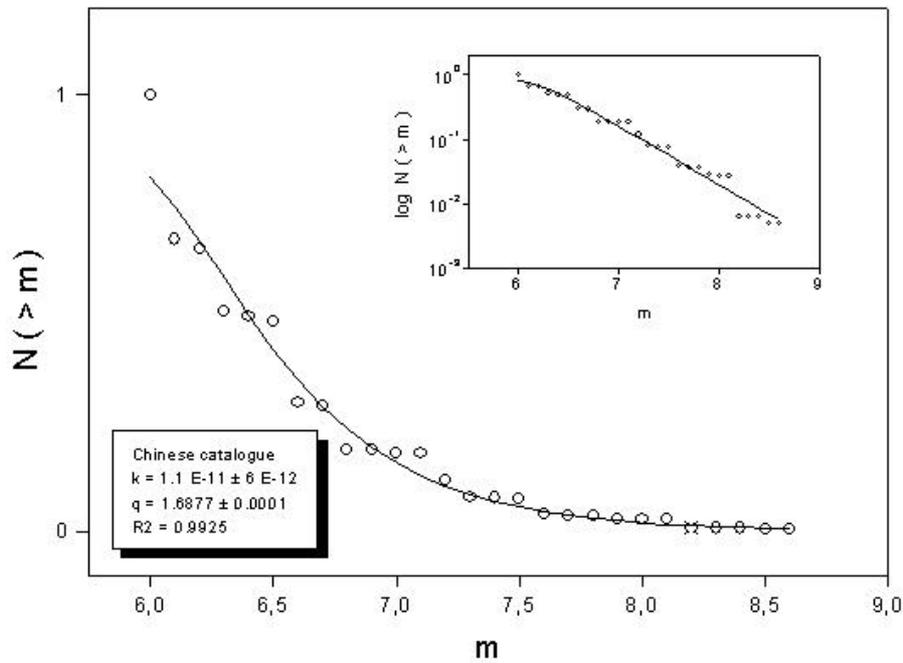

*Figure 6. Our model for large earthquakes (Chinese catalogue). The correlation factor is 0.992 and the value of q is 1.687.*

**Application of our model to a whole catalogue**

Two different catalogues are used in this section. First, a large catalogue from the United States Geological Survey including all the earthquakes in the California area, that is, all the San Andreas fault systems; the temporal period is from 1990 to the present time. More than 500000 earthquakes were processed and the results are in figure 7. As we can see, our formulation, based in Tsallis's statistics, describes all the earthquakes in the catalogue. The value of $q$ is 1.675 ± 0.001 and the correlation factor is 0.9985.

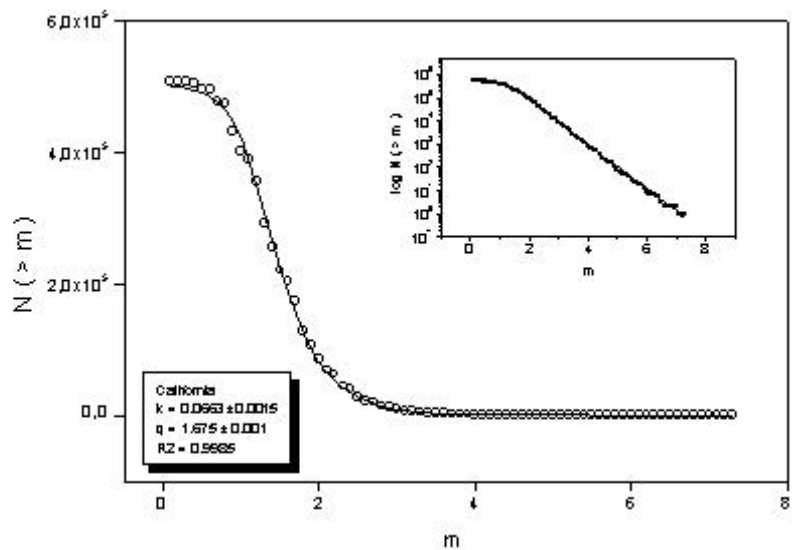

*Figure 7. Application of formula 9 for California earthquakes. The correlation factor is 0.999 and the value of q is 1.675.*

The second large catalogue is from the National Geographic Institute (Spain) and it has all the seismic data of the Iberian Peninsula (Spain). More than 10000 earthquakes are collected from 1970 to the present time. The results are shown in figure 8. Our formulation works well also with this data; the results are $q = 1.66 ± 0.01$ and the correlation factor 0.9931.

**CONCLUSIONS**

A functional dependence was obtained for the distribution of earthquakes produced by interactions in the space between the fault planes, starting from first principles, i.e., a non-extensive formulation of the maximum entropy principle (the Tsallis formulation). The Bolztmann entropy was also used for comparison to show its inadequacy. The active role of the material between the fault planes was revealed with this model. Non-extensivity is, as can be seen, determinant to obtain the energy distribution of earthquakes in a wide energy range. No "a priori " assumption about the fault profile or shape of the fragments was needed.

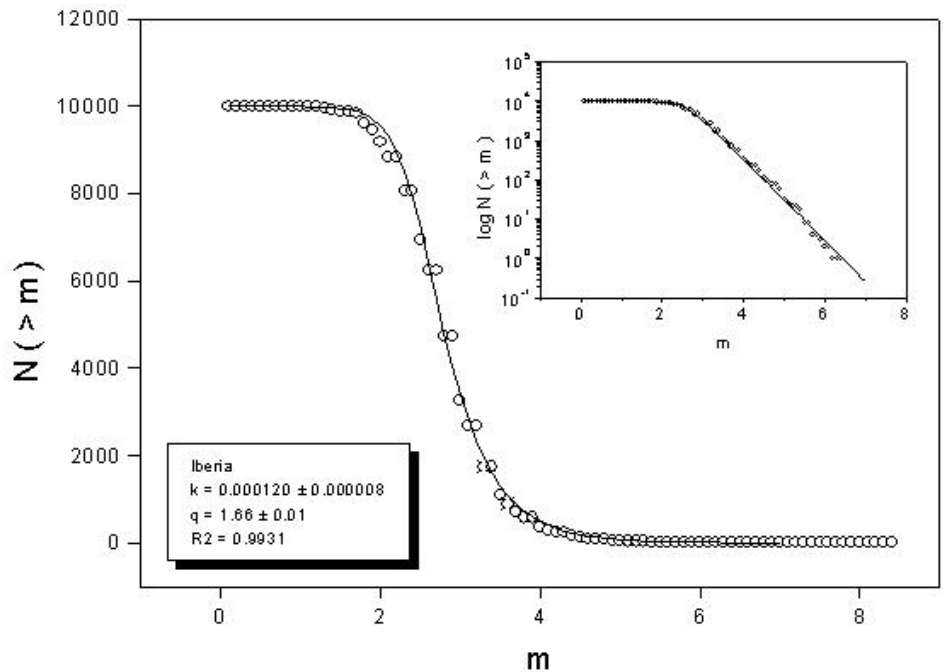

*Figure 8. Application of our model to the Iberian Peninsula earthquakes. The correlation factor is 0.993 and the value of q is 1.660.*

We performed two tests with small ($m < 5.0$) and large earthquakes ($m > 6.0$); the results lead to a similar value of $q$ (1.650 and 1.687 respectively). This means that our expression can fit both small and large earthquakes. After that we have used a whole catalogue of earthquakes to check the ability of our expression to fit the data. Results are good in both cases: the region of California lead us to $q = 1.675$ and the Iberian Peninsula region lead us to $q = 1.660$. It is very important to point out that our adjustment leads to a $q$ value equal approximately to 1.7. This is a very interesting result because it informs us about the scale of interactions in the gouge. It is known that $q \approx 1$ means short ranged spatial correlations and physical states close to equilibrium states (Boltzmann statistics). As $q$ increases, the physical state goes away from equilibrium states. A value of $q = 1.7$ means that the fault planes in the analyzed zone are not in equilibrium and more earthquakes can be expected.

Figures 5 to 8 show both the linear scale and the logarithmic scale representation for the cumulative number of earthquakes to highlight the nice agreement of our results with the observed data.

We think that all the results here exposed point in favor of a non extensive description of large scale correlated phenomena. The explanation of such a diverse collection of earthquake catalogues with the same formulation looks far from being casual, so that Tsallis entropy seems to be much more than a mathematical artifact.

It is very curious to observe the similarity in the value of the non-extensivity parameter "$q$" for all the used catalogues. This remains intriguing for us and we think that a more exhaustive study of the non-extensive statistics and its relation with earthquakes is needed to give a deeper interpretation of this value.

## ACKNOWLEDGEMENTS

This work was partially supported by the CICYT project REN2001-2418-C04-02/RIES, and the Alma Mater contest, University of Havana. One of us (O.S.) is grateful to the Department of Physics of the University of Almería for their kind hospitality.